\documentclass{osa-article}

\journal{osajournal}
\usepackage{multirow}
\usepackage{bm}
\usepackage{adjustbox}
\usepackage[utf8]{inputenc}

\articletype{article}
\begin{document}
\title{Silicon carbide soliton microcomb generation for narrow-grid optical communications}

\author{Jingwei Li,\authormark{1},Ruixuan Wang \authormark{1}, Haipeng Zheng \authormark{2}, Zhensheng Jia \authormark{*2}, and Qing Li\authormark{*1}}

\address{\authormark{1}Department of Electrical and Computer Engineering, Carnegie Mellon University, Pittsburgh, PA 15213, USA}
\address{\authormark{2} CableLabs, 858 Coal Creek Circle, Louisville, Colorado 80027, USA}
\email{\authormark{*}s.jia@cablelabs.com}
\email{\authormark{*}qingli2@andrew.cmu.edu} 


\begin{abstract}
A soliton microcomb can play a crucial role in narrow-grid optical communications by replacing many independently operated lasers in wavelength-division multiplexing systems. In this work, we designed and demonstrated power-efficient soliton microcombs with 100-GHz free spectral range in an integrated 4H-SiC platform for the first time. The combination of enabling technologies, including efficient fiber coupling (3 dB insertion loss), high-quality-factor microrings (intrinsic quality factors up to 5.7 million), and the employment of the Raman effect for adiabatic accessing of the soliton state, has enabled the demonstration of soliton pump power as low as 6 mW while supporting comb powers above -20 dBm per line near the pump wavelength. 
\end{abstract}
\noindent 
\section{Introduction}
A soliton microcomb source consists of multiple comb lines that are equally spaced and phase coherent with each other, which has revolutionized the traditional frequency comb technologies by reducing the device’s size, weight, and power (SWaP) while enabling system-level integration \cite{Diddams_comb_review1}. Soliton microcombs have found a wide range of applications such as frequency generation and synthesis \cite{Papp_comb_synthesizer, Comb_freq_XK, Wong_SiN_THz, Diddams_review_spectrum}, imaging and sensing \cite{Vahala_comb_imaging, Diddams_comb_sensing1, Comb_midIR_spectroscopy, Vahala19_comb_exoplanets}, light detection and ranging \cite{Vahala_comb_ranging, Kippenberg_comb_ranging}, parallel optical communication and computation \cite{Comb_communication, Comb_parallel_computation}, and quantum information processing \cite{Wong_SiN_quantum, Review_quantum_comb}. Several integrated photonic platforms, including silicon \cite{Gaeta_Si_comb, Gaeta_Si_dual_midIR}, silica \cite{Vahala_comb_silica, Xiao_comb_silica}, aluminum nitride \cite{Guo_comb_AlN, Tang_comb_AlN_ref}, silicon nitride \cite{Li_SiN_octave, Kippenberg_SiN_octave}, lithium niobate \cite{Loncar_LN_EOM, Lin_LN_soliton, Tang_LN_comb}, AlGaAs \cite{Pu_AlGAAs_comb, Bowers_comb_AlGaAs}, have shown promises for microcomb generation. 

In this work, we exploit an emergent integrated photonics platform, 4H-silicon-carbide-on-insulator (4H-SiCOI), for producing power-efficient soliton microcombs suitable for narrow-grid optical communications \cite{Noda_4HSiC_PhC, Vuckovic_4HSiC_nphoton, Ou_4HSiC_combQ}. With tailored properties, such a microcomb source could play a crucial role in network facilities, such as central offices or hubs, by replacing many independently operated lasers in wavelength-division multiplexing (WDM) systems, thereby significantly reducing the overall cost and power consumption of optical hardware \cite{Pu_comb_communication}. The 4H-SiCOI platform is selected mainly because of its strong Kerr nonlinearity, which is estimated to be four times of that of silicon nitride and thus allows a reduction of pump power by more than one order of magnitude (assuming other factors are the same) \cite{Li_SiC_Kerr_coeff}. Our device optimization has resulted in the first demonstration of single-soliton SiC microcomb with a free spectral range of 99.8 GHz, operable for pump powers as low as 6 mW. In addition, associated enabling technologies, such as efficient fiber coupling and the utilization of Raman-active resonances for adiabatically accessing the soliton state, have also been developed. 

\section{Device design and fabrication}
As illustrated in Fig.~1(a), the 4H-SiCOI integrated photonics platform consists of a thin-layer of 4H-SiC on top of lower-index cladding materials such as silicon dioxide. In this work, the SiC thickness is chosen to be 700 nm for the demonstration of microcombs tailored for narrow-grid optical communications. Using optimized nanofabrication, we obtain low-loss photonic devices, including waveguides and microresonators, with varied geometries (details see Ref.~\cite{Li_4HSiC_comb}). Figure 1(b) shows one such example, i.e., a 169-$\mu$m-radius SiC microring with a ring waveguide width of 3 $\mu$m. Its free spectral range (FSR) is expected to be near 100 GHz, a preferred choice for the alignment of the produced comb lines with the dense WDM International Telecommunication Union (ITU) grids. In Fig.~1(c), the simulated dispersion of the fundamental transverse-magnetic (TM$_{00}$) and transverse-electric (TE$_{00}$) modes is provided, where the integrated dispersion ($D_\text{int}$) is defined as: $D_\text{int} \equiv \omega_{\mu} - \omega_0 - D_1\mu$, with $\mu$ being the relative azimuthal order to the pump resonance (i.e., $\mu=0$ for the pump mode), $\omega_{\mu}$ being the corresponding resonance frequency, and $D_1/(2\pi)$ corresponding to the FSR. As can be seen, the TM$_{00}$ mode exhibits a relatively strong anomalous dispersion, with its dominant group velocity dispersion (GVD) computed to be near 400 ps/($\text{nm}\cdot\text{km}$). In contrast, the GVD of the TE$_{00}$ mode is still anomalous but much weaker. Compared to the TE$_{00}$ mode, the TM$_{00}$ resonance is likely to yield a higher average power per comb line, albeit at the cost of a smaller comb bandwidth. This prediction is confirmed by numerical simulations based on the Lugiato-Lefever equation (LLE) \cite{Coen_LLE}. For example, with an injection power of 5 mW to the TM$_{00}$ resonance, the power of comb lines adjacent to the pump is estimated to be larger than -20 dBm (10 $\mu$W, see Fig.~1(d)), which is sufficient for further amplification and data transmission in optical communications. 

\begin{figure}[ht]
\centering
\includegraphics[width=0.75\linewidth]{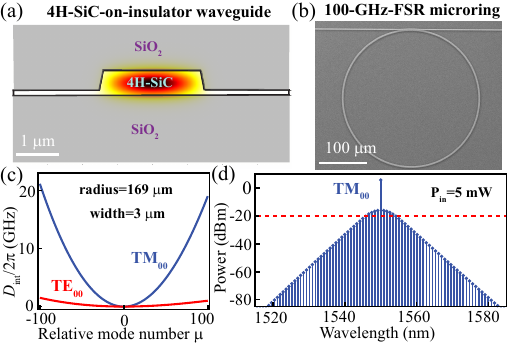}
\caption{(a) Illustration of a 4H-silicon-carbide-on-insulator (4H-SiCOI) waveguide (SiC thickness of 700 nm and a pedestal layer of 150 nm) and the mode profile of the fundamental transverse-magnetic (TM$_{00}$) mode. (b) Scanning electron micrograph of a 169-$\mu$m-radius SiC microring (ring width of $3\ \mu$m) with an expected free spectral range (FSR) of 100 GHz. (c) Simulated integrated dispersion ($D_\text{int}$) of the fundamental transverse-electric (TE$_{00}$) and TM$_{00}$ mode families in the 1550 nm band. (d) Simulated optical spectrum of the single-soliton microcomb corresponding to the TM$_{00}$ mode of a 100-GHz-FSR microring with an intrinsic (coupling) $Q$ of 5 (5) million. The Kerr nonlinear parameter and the on-chip power are assumed to be $3.2\ \text{W}^{-1}\text{m}^{-1}$ and 5 mW, respectively.}
\label{Fig1}
\end{figure}

\section{Fiber coupling and linear testing}
\begin{figure}[h]
\centering
\includegraphics[width=0.68\linewidth]{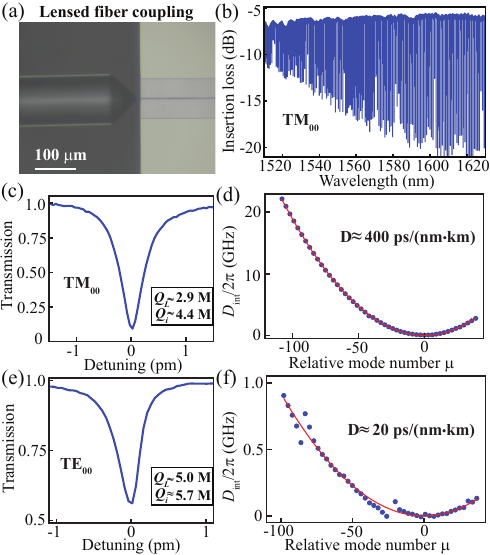}
\caption{(a) Optical micrograph of a lensed fiber coupled to an inverse taper implemented on the SiC chip. (b) Linear transmission scan of a 169-$\mu$m-radius SiC microring for the TM polarization, in which the TM$_{00}$ mode is dominantly excited. The total insertion loss is estimated to be within 6-7 dB range. (c) Representative resonance of the TM$_{00}$ mode with normalized transmission in the 1550 nm band, showing a loaded and intrinsic quality factors of $2.9$ million and $4.4$ million, respectively. (d) Dispersion characterization of the TM$_{00}$ mode families, displaying an approximate group velocity dispersion of 400 $\text{ps}/(\text{nm}\cdot\text{km})$.
(e) Representative resonance of the TE$_{00}$ mode with normalized transmission in the 1550 nm band, showing a loaded and intrinsic quality factor of $5.0$ million and $5.7$ million, respectively. (f) Dispersion characterization of the TE$_{00}$ mode families, displaying an approximate group velocity dispersion of 20 $\text{ps}/(\text{nm}\cdot\text{km}$).}
\label{Fig2}
\end{figure}

\noindent To minimize the coupling loss between the fiber and the SiC chip, we taper down the SiC waveguide width near the facets to increase the modal field overlap between the waveguide and a lensed fiber (see Fig.~2(a)). Numerical simulations point to a taper width of 250 nm to be best matched to a lensed fiber with a mode field diameter of $2.5\ \mu$m. In practice, the insertion loss is measured to be slightly higher than the theoretical value (2 dB, or $\approx 60 \%$ coupling efficiency), close to 3 dB ($\approx 50\%$ coupling efficiency) per facet instead (see Fig.~2(b)). In addition, we employ a pulley coupling scheme (see Fig.~1(b)) to selectively excite the desired resonant modes, which only allows the phase-matched resonances to be efficiently coupled. Take the transmission scan plotted in Fig.~2(b) as an example: it consists of a dominant excitation of the TM$_{00}$ mode families despite the multimode nature of the 3-$\mu$m-wide ring waveguide. For a coupling gap of 350 nm, the pulley-coupled TM$_{00}$ resonance is slightly under-coupled, displaying an intrinsic quality factor ($Q$) in the $3.5$-$4.5$ million range (see Fig.~2(c) for one instance). The coupling level becomes stronger as we move toward longer wavelengths, which is evidenced by the increased extinction ratio in the resonance scan (see Fig.~2(b)). On the other hand, the TE$_{00}$ modes are severely under-coupled, with the extracted intrinsic $Q$s in the $5$-$6$ million range (see Fig.~2(e) for one instance). Finally, we can characterize the resonator's dispersion by calibrating the recorded resonance frequencies with the aid of a wavemeter. The results provided in Fig.~2(d) and Fig.~2(f), corresponding to the respective dispersion of TM$_{00}$ and TE$_{00}$ mode families, are consistent with simulation data in Fig.~1(c). We also note that while the TM$_{00}$ resonances have shown minimum frequency deviation from the expected dispersion (Fig.~2(d)), several avoided mode crossings are observed for the TE$_{00}$ resonances (Fig.~2(f)). Such frequency shifts for the TE$_{00}$ mode may negatively impact the microcomb formation, which will be discussed later.  

\section{Nonlinear testing}
When increasing the pump power to the mW level, optical parametric oscillation (OPO) is observed for both TE$_{00}$ and TM$_{00}$ resonances due to their anomalous GVD and high optical $Q$s. We can estimate the threshold power of OPO as \cite{Vuckovic_4HSiC_soliton}
\begin{equation}
P_\text{th, OPO}\approx \frac{\pi n_g^2 V_{\text{eff}}}{4\lambda_p n_2}\cdot\frac{Q_{c,p}}{Q_{l,p}^3},
\label{Eq_Popo}
\end{equation}
where $n_g$ is the group index ($n_g\approx 2.8$); $V_\text{eff}$ is the effective mode volume ($V_\text{eff}\approx 1290\ {\mu m}^3$ for 169-$\mu$m-radius SiC microrings); $\lambda_p$ denotes pump wavelength; $n_2$ is the Kerr nonlinear index of 4H-SiC; and $Q_{c,p}$ and $Q_{l,p}$ are the coupling $Q$ and loaded $Q$ of the pump resonance, respectively. This relationship allows us to connect the Kerr nonlinear index of 4H-SiC to the observed OPO threshold power. For example, the OPO threshold of the TM$_{00}$ resonance near $1545.5$ nm is measured to be $2.7$ mW (see Fig.~3(a)), while its coupling and loaded $Q$ values are found to be $8.3$ million and $2.55$ million, respectively. Applying Eq.~\ref{Eq_Popo}, we infer a Kerr nonlinear index of $n_2\approx 9.5\times 10^{-19} \text{m}^2/\text{W}$. To account for the variations of the $Q$ factors among different azimuthal orders, we carry out the same measurement on other TM$_{00}$ and TE$_{00}$ resonances, and plot the extracted $n_2$ in Fig.~3(c). The statistically averaged Kerr nonlinear index index is found to be $n_2^{\text{TM}}\approx (9.4\pm 0.4) \times 10^{-19} \text{m}^2/\text{W}$ for the TM polarization and $n_2^{\text{TE}}\approx (8.8 \pm 0.4) \times 10^{-19} \text{m}^2/\text{W}$. Note that these numbers are consistent with the approach based on the four-wave mixing experiment carried out in a separate work \cite{Li_SiC_Kerr_coeff}.

\begin{figure}[ht]
\centering
\includegraphics[width=0.75\linewidth]{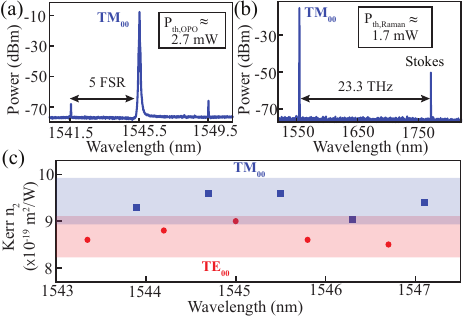}
\caption{(a) Optical spectrum showing the optical parametric oscillation (OPO) of the TM$_{00}$ resonance when pumped near $1545.5$ nm, with its OPO threshold power estimated to be near $2.7$ mW. (b) Optical spectrum showing stimulated Raman scattering of the TM$_{00}$ resonance when pumped near 1556 nm. The Raman threshold power is near $1.7$ mW, with the Stokes signal generated approximately $23.3$ THz away from the pump laser (corresponding to a Raman shift of $777\ \text{cm}^{-1}$). (c) Estimated Kerr nonlinear refractive index of 4H-SiC based on the measured OPO threshold power for various TM$_{00}$ and TE$_{00}$ resonances near 1550 nm.}
\label{Fig3}
\end{figure}

For high-$Q$ SiC microresonators, however, there is a competing Raman effect that exhibits a comparable threshold power as that of OPO \cite{Gaeta_Raman_competition}. In 4H-SiC, the dominant Raman frequency shift is around 777 $\text{cm}^{-1}$ (or $23.3$ THz) \cite{Ou_4HSiC_combQ}. The threshold power corresponding to the stimulated Raman scattering process is given by \cite{Li_SiC_Raman}
\begin{equation}
P_{\text{th, Raman}}\approx \frac{\pi^2n_g^2V_{\text{eff}}}{\lambda_p\lambda_s g_R}\cdot\frac{Q_{c,p}}{Q_{l,p}^2}\cdot\frac{1}{Q_{l,s}},
\label{Eq_Praman}
\end{equation}
where $\lambda_s$ and $Q_{l,s}$ denote the wavelength and loaded $Q$ of the Stokes resonance, respectively; and $g_R$ is the Raman gain coefficient which has a peak value of $g_R^\text{max}=(0.75 \pm 0.15) \ \text{cm}/\text{GW}$ in the 1550 nm band and an approximate full width at half maximum (FWHM) of ($120 \pm 30$) GHz \cite{Li_SiC_Raman}. One can understand the competition between the Raman and OPO effects in 4H-SiC by comparing their respective threshold powers as 
\begin{equation}
    \frac{P_\text{th,Raman}}{P_\text{th,OPO}}=\frac{4\pi n_2}{g_R\lambda_s}\cdot\frac{Q_{l,p}}{Q_{l,s}}\approx 0.85 \cdot\frac{g_R^{\text{max}}}{g_R}\cdot\frac{Q_{l,p}}{Q_{l,s}}.
\label{Eq_ratio}
\end{equation}
According to Eq.~\ref{Eq_ratio}, we deduce that if the FSR of the microresonator is much larger than the FWHM bandwidth of the Raman gain ($\approx 120$GHz), $g_R$, the actual Raman gain of the Stokes resonance, is likely to be much smaller than $g_R^\text{max}$. This is because unless there is accidental frequency matching, the Stokes resonance is typically misaligned to the center of the Raman gain spectrum. As a result, the Kerr effect has a high probability to dominate over the Raman process. On the other hand, when the FSR of the microresonator is reduced below the FWHM of the Raman gain, which is the case here, $g_R$ becomes comparable to $g_R^\text{max}$ (for 100-GHz FSR, the worst $g_R$ is $60\%$ of $g_R^\text{max}$). The dominant nonlinear effect, whether it is Raman or Kerr, hence depends on the exact frequency mismatching condition and the ratio of loaded $Q$s between the pump and Stokes resonances \cite{Lin_LN_soliton_tuning}. In fact, for the 169-$\mu$m-radius SiC microring, the TM$_{00}$ mode families are observed to exhibit two distinct behaviors. For resonances below 1550 nm, the OPO process appears first, such as the one shown in Fig.~3(a). In contrast, for certain resonances above 1550 nm, such as the 1556 nm resonance shown in Fig.~3(b), the Stokes Raman dominates over the Kerr effect with a lower power threshold. For this type of resonance, the Kerr effect is still strongly suppressed even after increasing the pump power above the OPO threshold (i.e., no microcomb generation) \cite{Gaeta_Raman_competition}. 

\section{Adiabatic accessing of soliton states}
\begin{figure}[ht]
\centering
\includegraphics[width=0.7\linewidth]{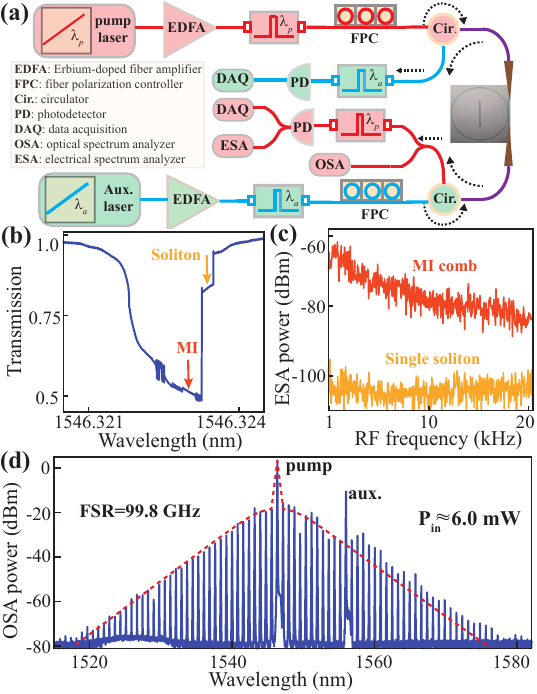}
\caption{(a) Experimental schematic for adiabatically accessing the single-soliton state with the aid of an auxiliary cooling laser. (b) Pump scan of the TM$_{00}$ resonance near the wavelength of $1546.3$ nm when the TM$_{00}$ resonance near 1556 nm is selected for the auxiliary laser. Both modulation instability (MI) and single-soliton states are observed for a pump power of 6 mW. (c) Recorded data from the electrical spectrum analyzer (ESA) corresponding to the MI (orange) and single-soliton (yellow) states. (d) Optical spectrum corresponding to the single-soliton state with a measured FSR of $99.8$ GHz (blue lines). The red dashed line is the simulated envelope of the single soliton state based on the LLE method.}
\label{Fig4}
\end{figure}

To generate soliton microcombs, we must overcome the thermo-optic bistability that prevents adiabatic accessing of the soliton state. While there are multiple approaches available, in this work we employ a cooling laser method by introducing an auxiliary laser to mitigate the thermal effect \cite{Wong_soliton_aux}. In general, the auxiliary laser is coupled to a resonant mode different from the pump to avoid competing comb generation (either with a different radial order or different polarization). Here, the competing Raman effect in 100-GHz-FSR microrings can be used to our advantage, as we can employ the TM$_{00}$ mode with the suppressed Kerr effect for the thermal cooling. For example, by pumping the TM$_{00}$ resonance near $1546.3$ nm (for which the Kerr effect dominates) while also coupling to the TM$_{00}$ resonance near $1556$ nm (for which the Stokes Raman dominates) in the counter-propagating direction (see the schematic in Fig.~4(a)), we have successfully observed soliton step in the pump transmission (Fig.~4(b)). The power of the auxiliary laser is typically adjusted between one to three times of the pump laser to maximize the soliton step. The dramatic change from the chaotic modulation-instability (MI) comb to the phase-coherent soliton state is reflected in the photodetected signal when it is connected to an electrical spectrum analyzer (Fig.~4(c)). Finally, the optical spectrum of the soliton state is recorded by slowly tuning the pump laser to the soliton step, confirming that it is indeed a single-soliton state for pump powers as low as 6 mW (Fig.~4(d)). The experimental data also agrees with the numerical simulation based on the LLE method, with the minor deviation attributed to stronger resonator-waveguide coupling at longer wavelengths (hence higher comb powers).

\begin{figure}[ht]
\centering
\includegraphics[width=0.9\linewidth]{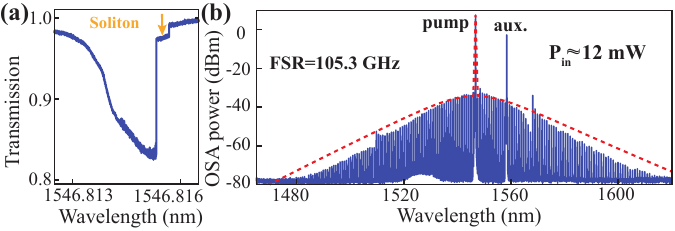}
\caption{(b) Pump scan of the TE$_{00}$ resonance near the wavelength of 1546nm when the TM$_{00}$ resonance near 1556 nm is selected for the auxiliary laser. A single soliton step is observed for an input power around 12 mW. (b) Optical spectrum corresponding to the single-soliton state. The red dashed line is the simulated envelope of the single soliton state based on the LLE method.}
\label{Fig5}
\end{figure}

For the TE$_{00}$ mode, we carry out the same procedure by employing one of the TM$_{00}$ resonances with the suppressed Kerr effect for thermal cooling. As a result, single soliton states can be adiabatically accessed for pump powers as low as 12 mW (see Fig.~5). While its comb bandwidth is verified to be broader than that of the TM$_{00}$ mode, the average power of the comb lines for the TE$_{00}$ mode is approximately 10 dB smaller. In addition, there are several avoided mode crossings appearing in the comb spectrum, which further lower the comb power and hinder the spectral expansion (see Fig.~5(b)).

Finally, it is worth comparing our results to Kerr soliton microcombs with similar FSRs reported in other material platforms. As can be seen in Table 1, the demonstrated soliton operation power (i.e., 6 mW) is among the lowest reported numbers in the literature (despite having relatively low $Q$s when compared to others). This is uniquely enabled by the stronger Kerr nonlinearity offered by 4H-SiC, which is approximately 4 times of that of SiN and 40 times of that of silica. In addition, we also note that our experiment represents the first demonstration of 100-GHz-FSR soliton microcombs in any SiC microresonators.

\begin{table}[ht]
\centering
\begin{tabular}{c c c c c}
\hline
\textbf{Material} & \textbf{FSR} (GHz) & $\bm Q_0$ (M) & \textbf{Soliton pump power} (mW) & \textbf{Reference}\\
\hline
\textbf{Silica} & $\bm{50.6}$ & $\approx 280$ & $\approx 3.0$ & S.~Zhang et.al.\cite{DelHaye_silica_comb}\\
\hline
\textbf{LN} & $\bm{44.84}$ & $\approx 4$ & 396 & Y.~He et.al.\cite{Lin_LN_soliton_tuning}\\
\hline
\multirow{2}{*}{\textbf{Si$_3$N$_4$}}& $\bm{99}$ & $\approx 15$ & $6.2$ & J.~Liu et.al. \cite{Kippenberg_SiN_100GHz_2018}\\
&$ \bm{99.79}$ & 8-14 & 5.0-7.4 & O.~B.~Helgason et.al.\cite{VT-Company_SiN_100GHz_2023} \\
\hline
\multirow{4}{*}{\textbf{4H-SiC}} & 350 & 5.6 & 2.3 & M.~A.~Guidry et.al. \cite{Vuckovic_4HSiC_soliton}\\
& 208 & 4.0& $\approx 39$ &C.~Wang et.al. \cite{OuXin_soliton}\\
& \textbf{99.8} & 4.4 (TM$_{00}$) & \textbf{6.0} & \textbf{This work}\\
&\textbf{105.3} & 5.7 (TE$_{00}$) & \textbf{12.0} & \textbf{This work}\\
\hline
\end{tabular}
 \caption{Selective list of Kerr soliton microcomb generation with microwave-rate FSRs (i.e., $<=100$ GHz) in various material platforms. For comparison within the 4H-SiCOI platform, We have also included experiments that have reported soliton states in SiC regardless of the FSR choice.}
\label{Table1}
\end{table}

\section{Conclusion}
In summary, we have demonstrated 100-GHz-FSR soliton microcombs in the 4H-SiCOI platform for the first time, which can be readily aligned to the DWDM ITU grid spacing. Through careful dispersion engineering and optimized nanofabrication, the on-chip pump power has been reduced to 6 mW while supporting comb powers above -20 dBm per line near the pump wavelength. In addition, our work has identified the potential limit from the Raman effect and how to use it for adiabatic accessing of the soliton state. As such, we believe the 4H-SiCOI platform can be highly competitive and promising for supporting low-power soliton generation with properties tailored for narrow-grid optical communications.
\begin{backmatter}
\bmsection{Funding}
This work was supported by NSF (2131162) and CableLabs University Outreach Program. 

\bmsection{Acknowledgments}
The authors would like to thank Dr.~Gregory Moille for helpful discussions on the soliton generation. The authors acknowledge the use of Bertucci Nanotechnology Laboratory at Carnegie Mellon University supported by grant BNL-78657879 and the Materials Characterization Facility supported by grant MCF-677785. R.~Wang also acknowledges the support of Tan Endowed Graduate Fellowship from CMU.  

\bmsection{Disclosures}  The authors declare no conflicts of interest.

\bmsection{Data Availability} Data underlying the results presented in this paper are not publicly available at this time but may be obtained from the authors upon reasonable request.

\end{backmatter}


\bibliography{SiC_Ref2}

\end{document}